\documentclass[runningheads]{llncs}

\usepackage[T1]{fontenc}
\usepackage{graphicx}
\usepackage{hyperref}
\usepackage{color}
\usepackage{listings}
\usepackage{booktabs}
\usepackage{array}
\usepackage{multirow}
\usepackage{amsmath}
\usepackage{microtype}
\usepackage[utf8]{inputenc}
\usepackage[most]{tcolorbox}

\newtcolorbox{takeaway}[1]{
  enhanced,
  colback=gray!9,
  colframe=gray!95,
  boxrule=0.5pt,
  arc=2pt,
  left=6pt, right=6pt, top=4pt, bottom=4pt,
  fontupper=\small,
  title={\small\textbf{#1}},
  attach boxed title to top left={yshift=-2pt, xshift=6pt},
  boxed title style={
    colback=gray!95,
    colframe=gray!95,
    arc=1pt,
    left=3pt, right=3pt, top=1pt, bottom=1pt
  }
}

\definecolor{lightgray}{gray}{0.95}
\definecolor{bordercolor}{gray}{0.75}

\lstdefinelanguage{Gherkin}{
  keywords={Feature, Scenario, Given, When, Then, And, But, Background, Scenario Outline, Examples},
  keywordstyle=\bfseries\color[rgb]{0.2,0.4,0.8},
  sensitive=false,
  comment=[l]{\#},
  commentstyle=\itshape\color{gray},
  stringstyle=\color[rgb]{0.0,0.5,0.0},
  morestring=[b]",
}
\usepackage{float}

\hypersetup{
    colorlinks=true,
    linkcolor=blue,   
    citecolor=blue,   
    urlcolor=blue     
}

\usepackage{orcidlink}
\begin{document}

\title{Epic-Organized vs.\ Requirement-Aligned Gherkin: An Empirical Evaluation of LLM-Based Acceptance Criteria Generation \thanks{This paper has been accepted for publication at the International Conference on Software Engineering of Emerging Technologies (SEET 2026). It will be published by Springer Nature in the conference proceedings}}

\titlerunning{Gherkin Representation Strategies: An Empirical Evaluation}

\author{Shahbaz Siddeeq\inst{1}\orcidlink{0009-0003-9030-8841} \and
Mateen Abbasi\inst{2}\orcidlink{0000-0001-8988-8816} \and
Jussi Rasku\inst{1}\orcidlink{0000-0002-4401-8013} \and
Zheying Zhang\inst{1}\orcidlink{0000-0002-6205-4210} \and
François Christophe\inst{2}\orcidlink{0000-0002-4337-3291} \and
Tommi Mikkonen\inst{2}\orcidlink{0000-0002-8540-9918} \and
Pekka Abrahamsson\inst{1}\orcidlink{0000-0002-4360-2226}}
\authorrunning{S. Siddeeq et al.}

\institute{Faculty of Information Technology and Communication Science, Tampere University, Tampere, Finland\\
\email{\{shahbaz.siddeeq,jussi.rasku,zheying.zhang,pekka.abrahamsson \}@tuni.fi} \and
Faculty of Information Technology, University of Jyväskylä, Jyväskylä, Finland \\
\email{\{mateen.a.abbasi,francois.m.christophe,tommi.j.mikkonen \}@jyu.fi}}

\maketitle

\begin{abstract}
Automated authoring of Gherkin Behavior-Driven Development (BDD) acceptance criteria remains a manual bottleneck in requirements engineering. This study investigates whether epic-organized LLM-generated Gherkin produces higher quality and coverage than requirement-aligned generation. We compare our \textit{Timeless} (an epic-organized LLM pipeline) approach against a naive large language model (LLM) baseline on four requirements documents (107 requirements) from the PURE dataset. Evaluation covers structural metrics, automated requirement coverage via TF-IDF and dense embeddings, and blind expert assessment by four researchers. In our evaluation, the JSON-constrained pipeline produced structurally valid scenarios across all generated outputs, while the zero-shot baseline achieved 99\% structural validity. Semantic coverage was comparable to the baseline, with Timeless achieving 94.3\% semantic Requirement Coverage Rate compared with 92.9\% for the baseline. TF-IDF produced lower coverage scores for the epic-organized output, suggesting that lexical metrics may miss coverage when scenarios paraphrase requirements at a higher level of abstraction. Expert raters prefer the epic-organized strategy on Correctness (4.61 vs 4.14), Executability (4.61 vs 4.07), and Completeness (4.31 vs 3.50). Overall, the results suggest that epic-organized generation can improve perceived Gherkin quality while maintaining comparable semantic coverage, although broader replication is needed before generalizing this finding.
\end{abstract}


\section{Introduction}
\label{sec:intro}

Software requirements are elicited primarily through stakeholder
meetings~\cite{Nuseibeh2000}, yet translating what has been agreed into formal, testable specifications remains a manual, time-consuming, and error-prone task~\cite{Kamsties1998,Lucassen2016}.
Behavior-Driven Development (BDD) addresses part of this by expressing acceptance criteria with a shared language. One such language is Gherkin, where senarios are described in \textit{Given}/\textit{When}/\textit{Then} format making them both
human-readable and executable by tools such as Cucumber~\cite{North2006,Solis2011,smart2023bdd,Cucumber2023}. In practice, however, Gherkin authoring remains manual: analysts write criteria after meetings from notes and memory, introducing three
problems~\cite{Lucassen2016,Binamungu2018}:
(1)~\textit{incompleteness}: edge cases discussed verbally are not captured;
(2)~\textit{delay}: criteria written days later lose tacit meeting context;
(3)~\textit{inconsistency}: criteria diverge in style across sessions.
A systematic mapping study found that automated tooling for Gherkin scenario authoring remains scarce, with teams consistently struggling to maintain well-formed scenarios as requirements evolve~\cite{BinamunguMaro2023}.

Large Language Models (LLMs) have made automated Gherkin generation from requirements feasible~\cite{OpenAI2023,Karpurapu2024}. However, an open question remains: does the \textit{representation strategy} (whether Gherkin use is organized around epics or individual requirements) affect structural quality, requirement coverage, and practitioner perception? While Gherkin is designed for epic- or story-level granularity~\cite{North2006,smart2023bdd}, existing automated tools generate scenarios at per-requirement granularity~\cite{Karpurapu2024}; comparing both strategies reveals the coverage--readability trade-off inherent to each approach.

We compare an epic-organized two-pass LLM pipeline (\textit{Timeless}) against a requirement-aligned zero-shot baseline across four PURE (Public Requirements) dataset documents (107 requirements in total) using structural metrics, dual-metric requirement coverage, and blind expert evaluation.

This paper makes the following contributions:
\begin{enumerate}
  \item An empirical comparison of epic-organized vs.\ requirement-aligned Gherkin generation (four PURE documents, 107 requirements) across structural quality, requirement coverage, and practitioner-perceived quality (four expert raters).
  \item A two-pass LLM prompting pipeline within \textit{Timeless} combining semantic epic synthesis with automated coverage-gap filling.
   \item Evidence that lexical-only coverage metrics can underestimate epic-organized Gherkin, motivating the use of complementary lexical and semantic coverage measures in future evaluations of LLM-generated BDD artifacts.
\end{enumerate}

Section~\ref{sec:background} reviews related work; Section~\ref{sec:approach} presents the generation approach; Section~\ref{sec:evaluation} reports results; Section~\ref{sec:discussion} discusses implications and limitations; Section~\ref{sec:threats} addresses threats to validity; Section~\ref{sec:conclusion} concludes.

\section{Background and Related Work}
\label{sec:background}

\subsection{Behavior-Driven Development, Gherkin, and the Authoring Challenge}

Behavior-Driven Development (BDD)~\cite{North2006,Solis2011} expresses software behavior through concrete scenarios in a shared language, bridging technical and non-technical stakeholders. Gherkin operationalizes this through a structured, human-readable syntax: a
\textit{Feature} block describes a product capability; each \textit{Scenario} specifies behavior via \textit{Given} (precondition), \textit{When} (action), and \textit{Then} (expected outcome) steps~\cite{North2006,smart2023bdd,Solis2011}. In BDD practice, related requirements are grouped into \textit{epics},
broad, cross-cutting capabilities that a user wants the system to provide. Each epic maps to one Gherkin Feature block, and the individual requirements within that epic inform the concrete scenarios: Epic $\to$ Feature block $\to$ Requirements $\to$ Scenarios~\cite{North2006,smart2023bdd}. This hierarchy is central to the paper's comparison: the epic-organized strategy
follows this design intent; the requirement-aligned baseline generates one scenario per requirement, bypassing the epic grouping level.

The primary barrier to BDD adoption is authoring effort: producing correct, complete Gherkin requires both domain knowledge and structural discipline, a combination difficult to sustain at scale~\cite{Binamungu2018}.

\subsection{Large Language Models for Acceptance Criteria and Gherkin Generation}

Recent LLMs have made automated Gherkin generation from raw requirements feasible~\cite{arora2024advancing,Ronanki2023}: GPT-4 matches human-analyst performance on requirements classification, and ChatGPT evaluates user story quality comparably to human reviewers with a consensus-based prompting strategy. Closest to our work, Karpurapu et al.~\cite{Karpurapu2024} conducted the most systematic evaluation to date, comparing GPT-4, GPT-3.5, and open-source models on BDD acceptance test generation from user stories: GPT-4 achieved the highest syntactic validity, but \emph{all} models struggled with negative-path scenario coverage. Rathnayake et al.~\cite{Rathnayake2026} confirmed high syntactic correctness
alongside variable semantic completeness across three LLM families.
Hassani et al.~\cite{hassani2026law} showed that Gherkin quality is sensitive to input \textit{representation format}, with iterative human review needed to recover quality lost to poorly structured inputs. Ferreira et al.~\cite{Ferreira2025} achieved a 95\% acceptance rate in an industrial two-step pipeline; however, their approach assumes already-structured user stories rather than raw SRS prose.

To the best of our knowledge, prior work has not jointly examined epic-organized Gherkin generation, automated coverage-gap filling, and comparison with a requirement-aligned LLM baseline using both lexical and semantic coverage measures. This study addresses that gap using SRS-derived functional requirements from the PURE dataset.

\section{Gherkin Generation Approach}
\label{sec:approach}

\subsection{Prompt Engineering for Gherkin Generation}
The Gherkin generation step receives the current requirements list and the generated epics. The system prompt instructs \texttt{gpt-4o-mini} to produce a JSON array in which each element specifies a \texttt{feature} name, an \textit{In~order~to / As~a / I~want} description, and a \texttt{scenarios} array of keyword-text step pairs, enforcing six structural constraints:

\begin{itemize}
  \item One \textit{Feature} block per epic (3–6 features total).
  \item Each Feature includes an \textit{In order to / As a / I want} description.
  \item Each Feature contains 2–4 Scenarios: at least one happy path and one negative or edge-case scenario.
  \item Steps use only \texttt{Given}, \texttt{When}, \texttt{Then},
        \texttt{And}, \texttt{But}.
  \item Steps are business-readable: no code, no technical jargon.
  \item Output is strictly valid JSON; no markdown fences or free prose.
\end{itemize}

These constraints address documented LLM failure modes~\cite{Karpurapu2024}: one Feature per epic follows Gherkin's design intent (Feature blocks describe capabilities at epic granularity~\cite{North2006,smart2023bdd}); mandating a negative or edge-case scenario counters the consistent under-generation of
failure paths observed across GPT-class models~\cite{Karpurapu2024}; and business-readable steps follow the BDD principle that criteria should communicate intent, not implementation~\cite{smart2023bdd,Solis2011}.

The JSON schema helps enforce the required output structure, while a deterministic validation step checks whether each scenario follows the expected Given–When–Then order. Together, these constraints reduce the chance of malformed Gherkin and explain the high structural validity observed in this evaluation. Listing~\ref{lst:gherkin} shows a representative Gherkin scenario rendered from the JSON-constrained output.

\begin{figure}[H]
\begin{lstlisting}[language=Gherkin, caption={Example Gherkin output generated by Timeless for a library management system.}, label={lst:gherkin}]
Feature: Book Reservation
  In order to reserve books in advance,
  As a registered library member,
  I want to place holds on unavailable books.

  Scenario: Successful reservation of an available book
    Given I am logged in as a library member
    When I search for "The Pragmatic Programmer"
    And the book has available copies
    Then I can reserve a copy
    And I receive a confirmation notification
\end{lstlisting}
\end{figure}

\subsection{The Timeless Platform}
\label{sec:system}

\begin{figure}[!tb]
\centering
\includegraphics[width=\textwidth, clip, trim=15 170 370 10]{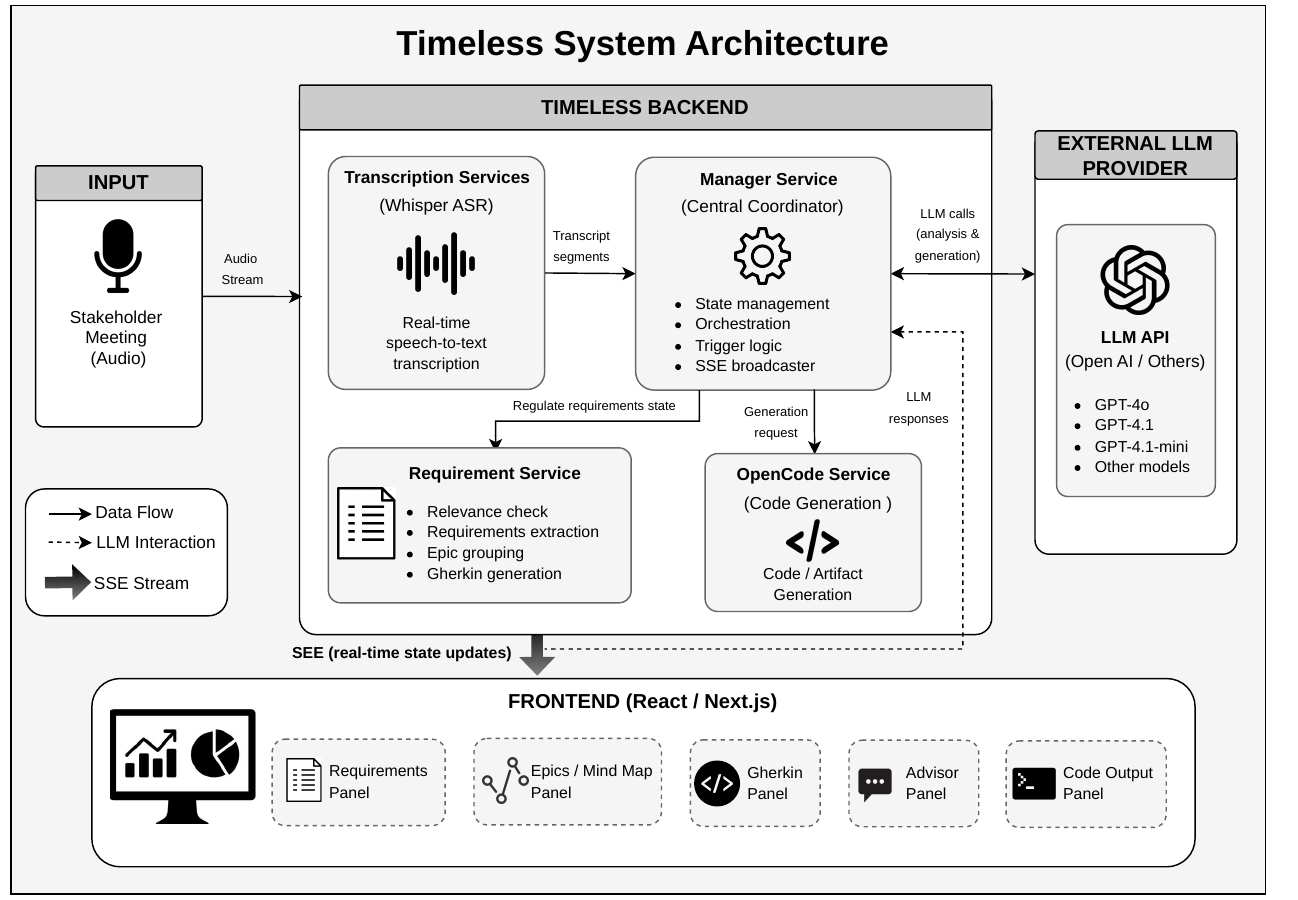}
\caption{Timeless system architecture. The Manager Service coordinates the
pipeline; SSE delivers real-time state updates to the React frontend.}
\label{fig:architecture}
\end{figure}

\textit{Timeless}~\cite{rasheed2024timeless} is a requirements engineering platform that transcribes meeting speech and extracts requirements in real time. Figure~\ref{fig:architecture} shows the microservice architecture: a Transcription Service captures and transcribes audio via OpenAI Whisper~\cite{Whisper2022}; a Requirements Service extracts and updates a requirements list from each transcription segment using an LLM; and a Manager Service coordinates both services, triggers downstream generation tasks (epics, mind map, Gherkin acceptance criteria), and streams state updates to the frontend via Server-Sent Events (SSE). The Manager
triggers generation when it detects new requirements through LLM-based relevance classification. This paper evaluates the Gherkin generation component.

\subsection{Pipeline Overview}

\begin{figure}[t]
\centering
\includegraphics[width=\textwidth, clip, trim=0 275 590 10]{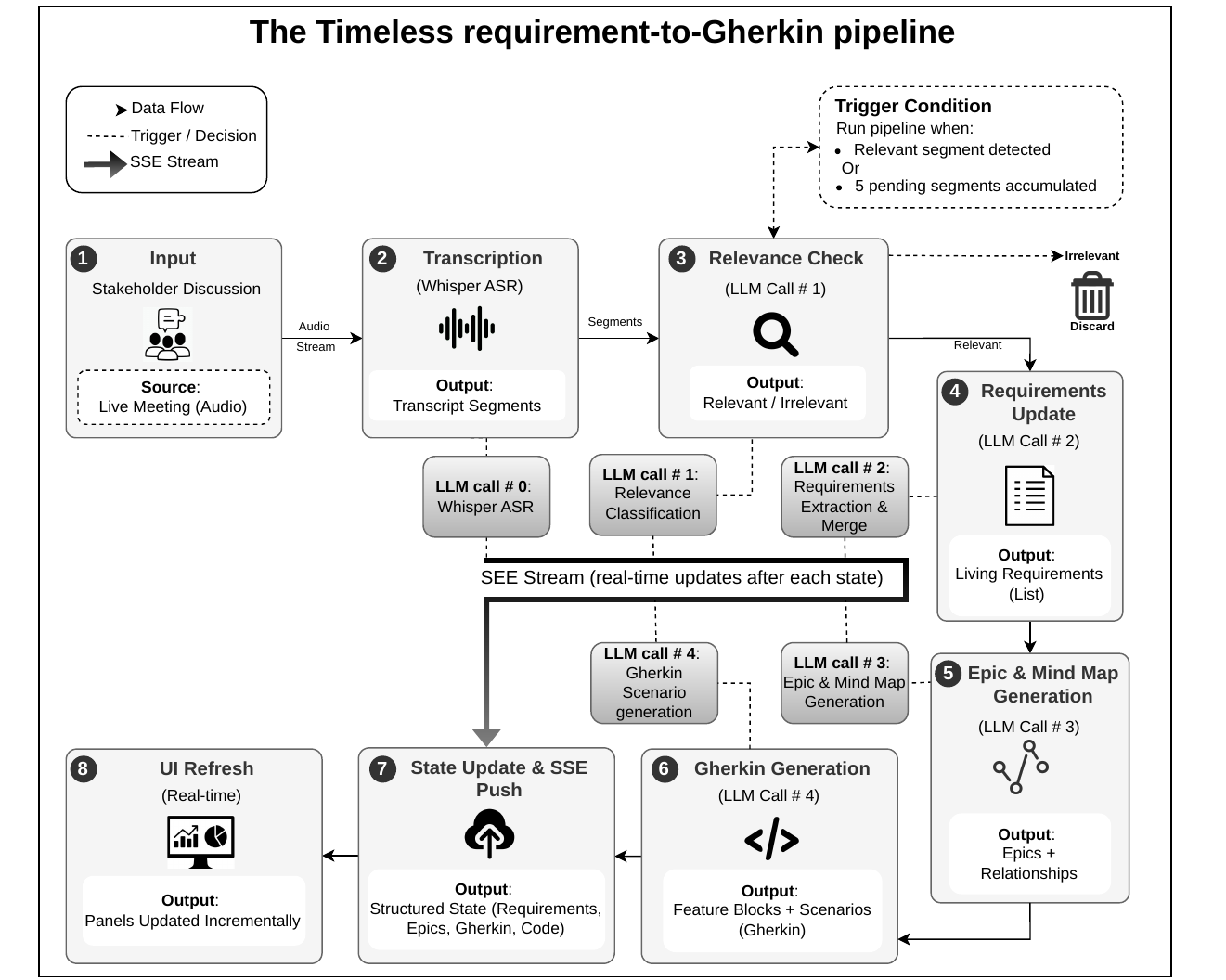}
\caption{The Timeless requirements-to-Gherkin pipeline. Each stage runs as an asynchronous LLM call; the SSE stream updates the User Interface (UI) as each stage completes.}
\label{fig:pipeline}
\end{figure}

Gherkin acceptance criteria are produced at the final stage of a four-step pipeline (Figure~\ref{fig:pipeline}): (1)~relevance classification (whether a transcription segment contains requirement-relevant content); (2)~requirements update (adding or revising items in the requirements list); (3)~epic and mind-map generation (grouping requirements into 3–6 epics); and (4)~Gherkin generation (producing Feature blocks with Given-When-Then scenarios for each
epic); Step 4 (Gherkin generation) runs after epics stabilize and is the focus of this evaluation.

\section{Evaluation}
\label{sec:evaluation}

\subsection{Research Questions}

We evaluate the approach through four Research Questions. The \textit{zero-shot LLM baseline} generates one Gherkin scenario per requirement in a single prompt, without epic grouping, JSON schema constraints, or gap-fill passes, representing the current state of automated LLM practice in the literature~\cite{Karpurapu2024,Rathnayake2026}.

\begin{itemize}
   \item \textbf{RQ1}: To what extent does an epic-organized, JSON-constrained generation pipeline produce syntactically valid and step-complete Gherkin scenarios compared with a zero-shot LLM baseline?
  \item \textbf{RQ2}: How do the requirement coverage characteristics (Requirement Coverage Rate [RCR] and Spurious Scenario Rate [SSR]) of epic-organized Gherkin compare to those of a zero-shot LLM baseline?
  \item \textbf{RQ3}: How does the epic-organized Gherkin \textit{representation} strategy (grouping requirements into semantically coherent epics, each represented as a Feature block, rather than one scenario per requirement) affect scenario quality as perceived by practitioners?
  \item \textbf{RQ4}: What are the practical trade-offs between epic-organized and requirement-aligned Gherkin generation, and what are the implications for BDD adoption in agile teams?
\end{itemize}

The RQs form a layered evaluation: structural correctness (RQ1) is a
prerequisite; coverage without sacrifice is the key comparison (RQ2, using dual lexical-semantic methods absent from prior BDD
studies~\cite{Karpurapu2024,Rathnayake2026,hassani2026law}); RQ3 adds
practitioner-perceived quality; RQ4 synthesizes actionable tradeoffs.

\subsection{Evaluation Design}

\subsubsection{Dataset.}
We used the PURE dataset~\cite{Ferrari2017} (79 SRS documents, Zenodo CC-BY 4.0). PURE's structured XML \texttt{<req>} elements allow consistent automated extraction of requirement statements, and the dataset has prior use as a Requirements Engineering (RE)--NLP benchmark~\cite{arora2024advancing,Ronanki2023}. We selected four documents covering distinct domains (security, e-commerce, law enforcement, networking), each with 20--35 functional requirements, exceeding the single-document scope of comparable
studies~\cite{Karpurapu2024,hassani2026law}. Table~\ref{tab:docs} summarizes the selection.

\begin{table}[t]
\centering
\caption{PURE documents selected for evaluation.}
\label{tab:docs}
\begin{tabular}{llcc}
\toprule
\textbf{System} & \textbf{Domain} & \textbf{Reqs used} & \textbf{PURE ID} \\
\midrule
KeePass Password Safe    & Security / Desktop     & 32  & keepass-2008   \\
Gamma-J Web Store        & E-Commerce / Web       & 30  & gamma-j-0000   \\
CCTNS Crime Tracking     & Gov. / Law Enforcement & 25  & cctns-0000     \\
Internet Peering Service & Networking / Services  & 20  & peering-2008   \\
\midrule
\textbf{Total}           &                        & \textbf{107} & \\
\bottomrule
\end{tabular}
\end{table}

\subsubsection{Gherkin Generation Procedure.}
For each selected document, requirements were extracted automatically using an XML parser targeting \texttt{<req>} elements. Duplicate sentences were removed. The resulting requirements list was passed to the Timeless Gherkin generation pipeline using the same LLM prompt and model (\texttt{gpt-4o-mini}, OpenAI) as the
production system. Generation latency (LLM call time) was recorded for each document. No manual editing of the generated output was performed.

\subsubsection{Baseline.}
The \textit{zero-shot baseline} uses the identical model (\texttt{gpt-4o-mini}) and requirements input as Timeless, but replaces the structured prompt with a minimal two-sentence instruction: \textit{``You are a software requirements expert. Generate Gherkin BDD acceptance criteria for the given requirements.''} No epic context, JSON schema, or BDD structural constraints are provided, isolating the contribution of prompt engineering from the underlying LLM capability.

\subsubsection{Metrics.}

We use two fully automated metric categories.

\paragraph{Structural quality (RQ1).}
Structural Validity (SV) measures whether \texttt{Given} precedes
\texttt{When} precedes \texttt{Then} in each scenario; out-of-order keywords cause parse-time rejection by BDD frameworks. Step Completeness (SC) measures whether each scenario contains at least one step of each mandatory type. Negative Coverage (NC) is the proportion of Feature blocks containing at least one negative or edge-case scenario, identified by keyword matching (``fail'', ``invalid'', ``error'', ``exceed''), a documented failure mode of LLM-generated Gherkin~\cite{Karpurapu2024}. Scenario Density is the ratio of generated scenarios to input requirements.

\paragraph{Requirement coverage (RQ2).}
Requirement \textit{coverage} here measures the proportion of input requirements addressed by at least one generated scenario; it is distinct from \textit{requirements traceability} as defined in IEEE 29148, which refers to bidirectional lifecycle artifact links.
Prior automated BDD evaluation studies report only lexical or human-assessed metrics, without automated requirement coverage
measurement~\cite{Karpurapu2024,Rathnayake2026,hassani2026law}; this study introduces dual measurement to isolate vocabulary bias from genuine coverage gaps. We compute TF-IDF cosine similarity ($\tau{=}0.20$, uni- and bigrams) for lexical coverage and OpenAI
\texttt{text-embedding-3-small} cosine similarity ($\tau{=}0.50$)~\cite{Reimers2019} for semantic coverage; dense embeddings provide a complementary view because they can capture some paraphrasing that TF-IDF may miss. Both methods compute RCR (proportion of requirements matched above threshold) and SSR (proportion of scenarios with no match above the selected threshold, indicating possible unsupported or weakly linked scenarios).

\subsection{Results}

\subsubsection{RQ1 — Structural Quality.}

Table~\ref{tab:auto} presents the automated structural metrics for
Timeless and the naive baseline across all four documents.

\begin{table}[t]
\centering
\caption{Automated structural quality metrics: Timeless and naive LLM
baseline. SV = Structural Validity; SC = Step Completeness (100\% in both systems; GPT-class capability floor, not a differentiating metric); NC = Negative Coverage; Dens = scenarios per requirement.}
\label{tab:auto}
\resizebox{\textwidth}{!}{%
\begin{tabular}{llcccccc}
\toprule
\textbf{Document} & \textbf{System} &
\textbf{SV\%} & \textbf{SC\%} & \textbf{NC\%} &
\textbf{Density} & \textbf{AvgSteps} & \textbf{Scen.} \\
\midrule
\multirow{2}{*}{KeePass}
  & Timeless        & 100.0 & 100.0 & 100.0 & 1.16 & 3.0 &  37 \\
  & Naive Baseline  & 100.0 & 100.0 & 100.0 & 0.75 & 3.0 &  24 \\
\midrule
\multirow{2}{*}{Gamma-J}
  & Timeless        & 100.0 & 100.0 &  66.7 & 0.87 & 3.5 &  26 \\
  & Naive Baseline  &  96.0 & 100.0 & 100.0 & 0.83 & 3.4 &  25 \\
\midrule
\multirow{2}{*}{CCTNS}
  & Timeless        & 100.0 & 100.0 &  72.7 & 1.36 & 3.1 &  34 \\
  & Naive Baseline  & 100.0 & 100.0 &  25.0 & 0.60 & 3.0 &  15 \\
\midrule
\multirow{2}{*}{Peering}
  & Timeless        & 100.0 & 100.0 &  28.6 & 1.10 & 4.0 &  22 \\
  & Naive Baseline  & 100.0 & 100.0 & 100.0 & 0.35 & 4.7 &   7 \\
\midrule
\multirow{2}{*}{\textbf{Average}}
  & \textbf{Timeless}       & \textbf{100.0} & \textbf{100.0} & \textbf{67.0} & \textbf{1.12} & \textbf{3.4} & \textbf{119} \\
  & \textbf{Naive Baseline} & \textbf{99.0}  & \textbf{100.0} & \textbf{81.2} & \textbf{0.60} & \textbf{3.5} & \textbf{71} \\
\bottomrule
\end{tabular}}
\end{table}

The JSON schema constraint eliminates step-ordering errors by design:
Timeless produces zero structurally invalid scenarios across all 119 generated, while the naive baseline records one mal-ordered scenario in the Gamma-J document (99\% SV). The distinction between the two systems is not the one-point margin but its \textit{nature}: Timeless’s perfect SV in this experiment reflects the effect of constrained generation and post-generation validation. Unlike the baseline, this approach explicitly checks the expected scenario structure before accepting the output. The baseline's near-perfect SV
is a \textit{statistical observation} that holds for this model and dataset but provides no guarantee for unseen inputs or different models.

Step completeness is 100\% in both systems (190 scenarios total), reflecting a GPT-class capability floor rather than a contribution of structured prompting.

The naive baseline achieves higher negative scenario coverage (81.2\%) than Timeless (67.0\%). The baseline groups requirements into broad features that each contain many scenarios, increasing the chance of including a negative case; Timeless's gap-fill pass creates focused features with fewer scenarios per feature.
Negative coverage correlates with scenario density, not prompt design: a feature with few scenarios rarely includes negative cases regardless of how it was generated.

\begin{takeaway}{Key Takeaway --- RQ1}
JSON schema enforcement makes structural validity an architectural property,
not a statistical outcome. The meaningful structural differentiator is
\textit{elaboration depth}: Timeless generates 1.12 scenarios per requirement
versus 0.60 for the zero-shot baseline, nearly double the coverage per input.
\end{takeaway}

\subsubsection{RQ2 — Requirement Coverage.}

We apply lexical (TF-IDF) and semantic (dense embedding) coverage analyses to isolate epic-level abstraction from genuine coverage gaps.

\begin{table}[t]
\centering
\caption{Requirement coverage: Timeless and naive LLM baseline.
TF-IDF ($\tau{=}0.20$) and semantic embeddings ($\tau{=}0.50$).
RCR = Requirement Coverage Rate; SSR = Spurious Scenario Rate.}
\label{tab:traceability}
\resizebox{\textwidth}{!}{%
\begin{tabular}{llcc|cc|cc}
\toprule
& & & & \multicolumn{2}{c|}{\textbf{Lexical (TF-IDF)}} &
         \multicolumn{2}{c}{\textbf{Semantic (Embeddings)}} \\
\textbf{Document} & \textbf{System} & \textbf{Reqs} & \textbf{Scen.} &
\textbf{RCR\%} & \textbf{SSR\%} &
\textbf{RCR\%} & \textbf{SSR\%} \\
\midrule
\multirow{2}{*}{KeePass}
  & Timeless         & 32 &  37 & 68.8 & 35.9 & 84.4 & 10.3 \\
  & Naive Baseline   & 32 &  24 & 75.0 & 12.5 & 90.6 &  0.0 \\
\midrule
\multirow{2}{*}{Gamma-J}
  & Timeless         & 30 &  26 & 80.0 & 19.2 &  96.7 &  0.0 \\
  & Naive Baseline   & 30 &  25 & 80.0 &  4.0 & 100.0 &  0.0 \\
\midrule
\multirow{2}{*}{CCTNS}
  & Timeless         & 25 &  34 & 44.0 & 67.6 &  96.0 & 20.6 \\
  & Naive Baseline   & 25 &  15 & 64.0 &  0.0 &  96.0 &  0.0 \\
\midrule
\multirow{2}{*}{Peering}
  & Timeless         & 20 &  22 & 95.0 &  9.1 & 100.0 &  9.1 \\
  & Naive Baseline   & 20 &   7 & 85.0 &  0.0 &  85.0 &  0.0 \\
\midrule
\multirow{2}{*}{\textbf{Average}}
  & \textbf{Timeless}       & & & \textbf{72.0} & \textbf{33.0} & \textbf{94.3} & \textbf{10.0} \\
  & \textbf{Naive Baseline} & & & \textbf{76.0} & \textbf{4.1}  & \textbf{92.9} & \textbf{0.0} \\
\bottomrule
\end{tabular}}
\end{table}

\paragraph{Two-pass coverage.}
The two-pass design achieves \textbf{94.3\%} semantic RCR, which is comparable to the naive baseline at (92.9\%), while TF-IDF RCR is only 72.0\%. Three documents reach $\geq$96\% semantic RCR (100\% on Internet Peering), driven by the explicit coverage mandate and second-pass gap-fill.

\paragraph{Lexical vs.\ semantic gap (TF-IDF bias).}
The 22.3~pp lexical-semantic gap (72.0\% vs.\ 94.3\%) confirms TF-IDF
underestimates abstracted Gherkin: Timeless scenarios paraphrase requirements at epic granularity, whereas the baseline echoes requirement wording directly (baseline gap: +16.9~pp). For this reason, lexical coverage should be interpreted alongside semantic coverage when evaluating structured Gherkin-generation pipelines.

\paragraph{Spurious Scenario Rate.}
The TF-IDF SSR for Timeless averages 33.0\%, particularly elevated for CCTNS (67.6\%). This is expected: the second-pass gap-fill generates scenario clusters that address \textit{groups} of uncovered requirements rather than individual ones, so no single scenario achieves lexical overlap above threshold with any individual requirement. Semantically, SSR is only 10.0\%, confirming these scenarios are grounded in requirements; the elevated TF-IDF SSR is a measurement artifact.

\begin{takeaway}{Key Takeaway --- RQ2}
Epic-organized generation achieves \textbf{94.3\%} semantic RCR,
equivalent to the requirement-aligned baseline (92.9\%), with no coverage penalty from epic-level grouping. The apparent lexical deficit (72.0\% vs 76.0\% TF-IDF RCR) is a measurement artifact: TF-IDF cannot match paraphrased epic-level scenarios to atomic requirement wording, underestimating Timeless by 22~pp. Evaluating structured Gherkin pipelines requires dual-metric coverage measurement (lexical \textit{and} semantic); lexical-only metrics systematically favor less abstracted systems.
\end{takeaway}

\subsubsection{Generation Latency.}

\begin{figure}[t]
\centering
\includegraphics[width=1.00\textwidth]{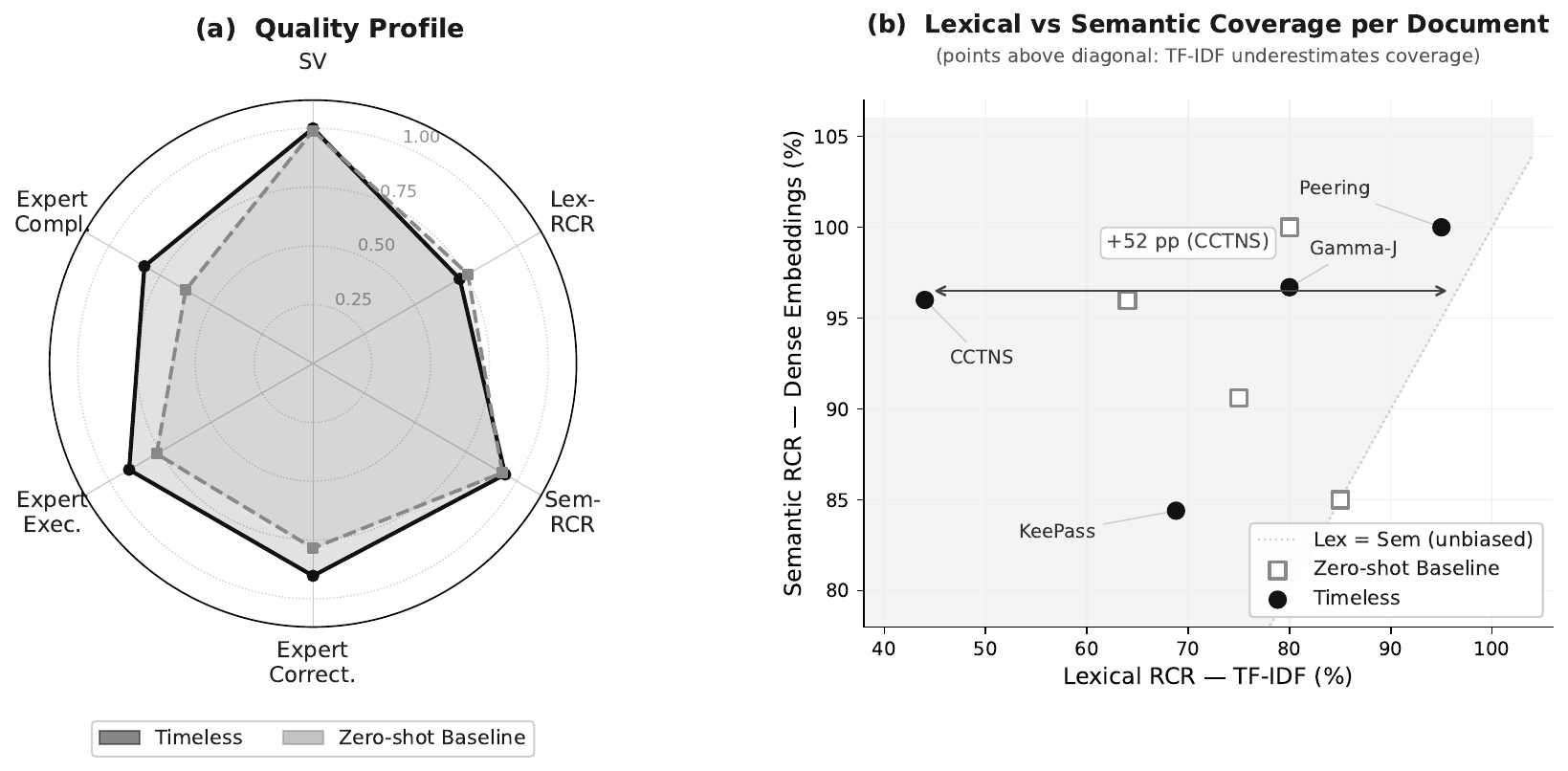}
\caption{Results overview. \textbf{(a)} Quality profile: six headline metrics normalized to [0,\,1] (higher = better for all axes). Timeless (solid, filled) and the zero-shot baseline (dashed, open markers) are compared across structural validity, lexical and semantic coverage, and the three expert evaluation dimensions.
\textbf{(b)} Lexical vs semantic RCR per document; filled circles = Timeless, open squares = baseline. Points above the diagonal (\textit{Lex\,=\,Sem}) indicate TF-IDF underestimation of coverage. The CCTNS document shows the largest gap: Timeless achieves 44\% lexical but 96\% semantic RCR (+52\,pp), confirming that TF-IDF systematically underestimates abstracted, epic-organized Gherkin.}
\label{fig:results}
\end{figure}

The Timeless two-pass pipeline averages \textbf{58.9 seconds} per document (\texttt{gpt-4o-mini}), compared to 21.5 seconds for the naive baseline. The additional latency reflects the second-pass gap-fill call; both pipelines complete in under 65 seconds, so acceptance criteria are available for practitioner review within the meeting in which requirements are elicited. This latency tradeoff, together with the NC and SSR tradeoffs identified above, informs the answer to RQ4 below. Figure~\ref{fig:results} summarizes the key automated and expert metrics across both systems.

\subsection{Human Expert Evaluation (RQ3)}

\subsubsection{Protocol.}
To evaluate the representational quality of generated Gherkin from a
practitioner perspective, we conducted a pre-registered blind expert
evaluation~\cite{Lucassen2016} archived on OSF\footnote{\url{https://osf.io/9apdf}}. Four SE researchers from four independent organizations served as evaluators (all PhD researchers; all with RE and/or BDD experience). Each received a blinded survey (Microsoft Forms) with four sections, one per document, presenting the full requirements list alongside System~A and System~B Gherkin outputs, randomly assigned to Timeless and the naive baseline (fixed seed; decode key archived on OSF post-collection). Each system was rated on three 1–5 Likert dimensions: \textbf{Correctness} (are Given/When/Then steps logically valid?),
\textbf{Executability} (can a developer implement step definitions directly?), and \textbf{Completeness} (do scenarios collectively cover the requirements?). Correctness and Executability are per-scenario; Completeness is per-document. Inter-rater agreement is Fleiss' $\kappa$. Pre-registered hypotheses: \textbf{H1}: Timeless scores higher on Executability (behavior-level steps are more directly implementable); \textbf{H2}: Timeless scores higher on
Correctness (JSON schema enforces structure); \textbf{H3}: the baseline scores higher on Completeness (higher requirement-aligned lexical coverage).

\subsubsection{Results.}

\begin{table}[t]
\centering
\caption{Human expert evaluation results (mean scores, 4 raters).
Correctness and Executability are per-scenario means; Completeness
is per-document mean (1 rating per rater). $\kappa$ = Fleiss' kappa for Correctness ratings pooled across both systems per document; near-zero values reflect systematic rater scale-usage differences, not directional disagreement.}
\label{tab:expert}
\resizebox{\textwidth}{!}{%
\begin{tabular}{llccc|c}
\toprule
\textbf{Document} & \textbf{System} &
\textbf{Correctness} & \textbf{Executability} & \textbf{Completeness} &
\textbf{$\kappa$} \\
\midrule
\multirow{2}{*}{KeePass}
  & Timeless       & 4.38 & 4.36 & 4.25 & \multirow{2}{*}{$-$0.08} \\
  & Naive Baseline & 4.16 & 4.11 & 3.50 & \\
\midrule
\multirow{2}{*}{Gamma-J}
  & Timeless       & 4.73 & 4.71 & 4.25 & \multirow{2}{*}{0.00} \\
  & Naive Baseline & 4.23 & 4.26 & 4.00 & \\
\midrule
\multirow{2}{*}{CCTNS}
  & Timeless       & 4.74 & 4.75 & 4.25 & \multirow{2}{*}{0.00} \\
  & Naive Baseline & 4.37 & 4.22 & 3.50 & \\
\midrule
\multirow{2}{*}{Peering}
  & Timeless       & 4.61 & 4.60 & 4.50 & \multirow{2}{*}{0.03} \\
  & Naive Baseline & 3.82 & 3.68 & 3.00 & \\
\midrule
\multirow{2}{*}{\textbf{Average}}
  & \textbf{Timeless}       & \textbf{4.61} & \textbf{4.61} & \textbf{4.31} & \multirow{2}{*}{\textbf{--}} \\
  & \textbf{Naive Baseline} & \textbf{4.14} & \textbf{4.07} & \textbf{3.50} & \\
\bottomrule
\end{tabular}}
\end{table}

Timeless scores higher than the naive baseline on all three dimensions across all four documents (Table~\ref{tab:expert}).
Mean Correctness is 4.61 vs 4.14 and mean Executability is 4.61 vs 4.07, supporting H1 and H2. The largest margin appears in Internet Peering (Correctness $\Delta{=}{+}0.79$, Executability $\Delta{=}{+}0.92$), where the baseline's flat, requirement-echoing scenarios diverge most from executable behavior descriptions.

Contrary to H3, Timeless scores \textit{higher} on Completeness (4.31 vs 3.50, $\Delta{=}{+}0.81$). Raters perceived epic-organized Gherkin as more complete than the requirement-aligned output, an unexpected result we attribute to Timeless's higher scenario density (1.12 vs 0.60 per requirement) and the explicit second-pass coverage mandate. The baseline's requirement-aligned structure, despite echoing requirement wording directly, does not translate into perceived coverage at the Feature-block level.

Fleiss' $\kappa$ is near zero across all documents ($\kappa = -0.08$ to $0.03$), reflecting systematic rater scale-usage differences: one rater consistently used the lower end of the scale while three used the upper range. Despite this, the directional ranking (Timeless $>$ Baseline) holds for 15 of 16 rater-document pairs on Correctness and Executability, and for all 16 on Completeness. All evaluation materials and analysis scripts are publicly available at \url{https://osf.io/9apdf}.

\begin{takeaway}{Key Takeaway --- RQ3}
Expert evaluation confirms epic-organized Gherkin is preferred across all three practitioner dimensions. \textbf{H1} (Executability: 4.61 vs 4.07) and \textbf{H2} (Correctness: 4.61 vs 4.14) are supported. \textbf{H3 is rejected}: Timeless also outscores the baseline on Completeness (4.31 vs 3.50, $\Delta{=}{+}0.81$), contrary to the pre-registered hypothesis; practitioners perceive epic-organized Gherkin as \textit{more} complete, attributable to higher scenario density and the explicit second-pass coverage mandate.
\end{takeaway}

\begin{takeaway}{Key Takeaway --- RQ4}
Teams prioritizing BDD maintainability and readability should adopt
epic-organized generation. Those requiring strict per-requirement lexical coverage tracking can augment with a one-to-one pass. The 37-second latency overhead is acceptable for meeting-time delivery of acceptance criteria.
\end{takeaway}

\section{Discussion}
\label{sec:discussion}

\paragraph{For researchers.}
The 22~pp gap between TF-IDF RCR (72.0\%) and semantic RCR (94.3\%) shows that lexical metrics undercount coverage of abstracted, epic-organized Gherkin. Studies that report only TF-IDF or BLEU-type coverage metrics for BDD
generation~\cite{Karpurapu2024,Rathnayake2026,hassani2026law} should be interpreted with this bias in mind: vocabulary mismatch, not missing coverage, drives the gap. More fundamentally, this is a construct validity concern: comparing a structured pipeline against a requirement-echoing baseline using only TF-IDF measures vocabulary retention, not semantic coverage, and systematically favors the less abstracted system. Future evaluations should report the lexical-semantic gap alongside both metrics; a large gap signals representation abstraction, not low coverage.

\paragraph{For practitioners.}
A 59-second average generation time places scenario authoring within the requirements meeting itself; acceptance criteria are available for review before the meeting closes, not drafted separately the next day. JSON schema enforcement eliminates a concrete CI failure mode: malformed Gherkin breaks a build at feature-file loading, before a single test runs; schema-constrained output passes that check by construction, removing a manual validation step from the developer's workflow. Epic-organized scenarios match the Feature/Scenario block structure that Cucumber and SpecFlow expect~\cite{North2006,smart2023bdd}, so output can be pasted directly into a \texttt{.feature} file and executed without restructuring; practitioners rated this output higher on Correctness (4.61 vs 4.14), Executability (4.61 vs 4.07), and Completeness (4.31 vs 3.50)~\cite{North2006,smart2023bdd}. Teams with per-requirement traceability obligations for compliance or audit can run an additional single-pass generation to produce requirement-aligned
output alongside the epic-organized structure, preserving both a readable BDD artifact and a per-requirement audit trail.

\paragraph{Limitations.}
The evaluation uses four documents from the PURE dataset (107 requirements); while consistent with comparable studies~\cite{Karpurapu2024,hassani2026law}, this limits confidence in the observed trends. All documents contain 20--35 functional requirements; it is not known whether the epic-grouping strategy
scales to larger SRS documents with 100+ requirements, where finer feature granularity may be needed.

\section{Threats to Validity}
\label{sec:threats}

\paragraph{Construct validity.}
Provider coupling between \texttt{gpt-4o-mini} (generation) and
\texttt{text-embedding-3-small} (embedding) means the semantic RCR score may be inflated if both models share similar internal representations; replication with Sentence-BERT~\cite{Reimers2019} would isolate this effect. Structural metrics do not capture readability, domain accuracy, or test precision~\cite{Lucassen2016}, which the expert evaluation addresses.

\paragraph{Internal validity.}
The zero-shot baseline isolates structured prompting's contribution; a chain-of-thought or few-shot baseline would narrow the NC and expert-rating gaps, though the architectural SV guarantee would remain. Results should be read as structured vs.\ unstructured prompting, not as the upper bound of LLM capability.

\paragraph{External validity.}
Four documents yield four data points per metric, sufficient to show consistent trends but not enough for statistical generalization. Documents were selected to cover distinct domains (security, e-commerce, law enforcement, networking), following the sampling strategy of comparable LLM-based RE evaluations~\cite{Ronanki2023,arora2024advancing}.

\paragraph{Conclusion validity.}
All results are from a single run per document with \texttt{gpt-4o-mini} (OpenAI, April 2026); LLM non-determinism means repeated runs may produce different scenario counts and content. Structural guarantees depend on the JSON-constrained prompting strategy and must be re-validated for each new model or version.

\section{Conclusion and Future Work}
\label{sec:conclusion}

Gherkin scenario authoring requires domain knowledge, structural discipline, and full requirement coverage, a combination difficult to sustain manually at scale. This paper evaluates a two-pass LLM prompting approach that organizes requirements into semantically coherent epics before generating JSON-constrained Gherkin output, compared against a zero-shot baseline across four PURE SRS documents (107 requirements).

JSON-constrained generation achieves 100\% structural validity by architectural design. Semantic coverage reaches 94.3\%, matching the zero-shot baseline (92.9\%), while TF-IDF underestimates abstracted Gherkin by 22~pp, a systematic bias lexical-only studies cannot detect. Expert evaluation confirms that epic-organized output scores higher on all three quality dimensions, including Completeness where the pre-registered hypothesis predicted the reverse.

Three directions follow from these findings. First, extending the evaluation to $n \geq 10$ PURE documents and live transcription inputs will test whether the coverage and quality trends hold across a wider requirement dataset (Section~\ref{sec:threats}, external validity). Second, comparing the zero-shot baseline with chain-of-thought and few-shot prompts will separate the contribution of epic organization from the benefit of richer prompting alone (Section~\ref{sec:threats}, internal validity). Third, replication
with open-source embedding models such as Sentence BERT~\cite{Reimers2019} will isolate the provider coupling concern in the semantic coverage measure.

The results suggest that epic-organized, schema-constrained generation is a promising direction for automated BDD support. Further evaluation with stronger baselines, more documents, repeated model runs, live meeting inputs, and manually validated traceability data is needed before the findings can be generalized.

\section*{Declaration of AI Assistance}
During the preparation of this manuscript, the authors used ChatGPT to assist with grammar refinement, sentence restructuring, and formatting improvements. Following the use of this tool, the authors carefully reviewed and revised the content and assume full responsibility for the final version of the publication.

\section*{Acknowledgment}
This work has been supported by FAST, the Finnish Software Engineering Doctoral Research Network, funded by the Ministry of Education and Culture, Finland and ANSE (AI Native Software Engineering) (1822/31/2025).

\bibliographystyle{splncs04}
\bibliography{references}

\end{document}